\documentclass[twocolumn,showpacs,superscriptaddress]{revtex4}
\usepackage{epsfig}
\usepackage{amssymb}
\usepackage{amsmath}
\usepackage{amsbsy}
\usepackage{color}
\usepackage{ulem}
\usepackage{epstopdf}

\begin{document}
\title{Statistics and Geometrical Picture of Ring Polymer Melts and Solutions}
\author{Takahiro Sakaue}
\email{sakaue@phys.kyushu-u.ac.jp}
\affiliation{Department of Physics, Kyushu University 33, Fukuoka 812-8581, Japan}
\affiliation{PRESTO, Japan Science and Technology Agency (JST), 4-1-8 Honcho Kawaguchi, Saitama 332-0012, Japan}

\begin{abstract}
We present a detailed account of a recently proposed phenomenological theory for noncatenated ring polymer melts (Phys. Rev. Lett. {\bf 106}, 167802 (2011)). A basic assumption lies in the implementation of the noncatenation constraint via the effective excluded-volume effect, from which a geometrical picture of melts emerges. The result captures many of the salient features observed so far, including (i) the overall spacial size of rings, (ii) the coordinate number, i.e., the number of rings surrounding  a given ring, (iii) the topological length scale as a function of the molecular weight and (iv) the effect of the chain stiffness and concentration. We also suggest a geometrical interpretation of the topological length scale, which may shed some light on the entanglement concept in polymeric systems.
\end{abstract}


\maketitle

\section{Introduction}
\label{Introduction}
The phenomenon of entanglement is ubiquitous in systems of long flexible ropes and, once formed, it often has crucial consequences on the subsequent operations, as we know from our daily experience~\cite{PNAS_2006}.
It also has a fundamental importance in molecular systems, dictating the dynamics and rheology of dense linear polymer solutions of high molecular weight~\cite{deGennes,Doi_Edwards,Rubinstein_Colby}.
Entanglement originates, in any case, from the topological constraints imposed by the non-crossability of the chain. 
In the history of polymer science, predicting the macroscopic viscoelasticity of the entangled polymer solution from the molecular theory has been one of the major goals.
Our current understanding on this issue owes to the introduction of the following two theoretical models.
The tube model, originally proposed in the theory of rubber elasticity, assumes that the topological constraint imposed by the non-crossability of the chain effectively confines each chain inside a tube-like region, thereby reducing the original many-chain problem to single-chain problem in the ``mean-field"~\cite{Edwards_1967}.
The reptation model then describes the overall motion of the chain under the temporal topological constraint along the primitive path, where the presence of the chain end is essential for the disengaging from the original tube and the subsequent tube renewal~\cite{deGennes_1971}.

In this context, an interesting question arises in the system of ring polymers, also called circular or cyclic polymers~\cite{C-D,McLeish}. Here, by definition, there are no chain ends, and the topology of the system is frozen at the stage of the sample synthesis, the consequence of which is pronounced even for static properties. To emphasize the point, let us recall the basic static properties of linear polymers~\cite{deGennes,Rubinstein_Colby,Grosberg_Khokhlov}. In dilute solutions, their conformation can be characterized as a self-avoiding walk with the overall chain dimension $R_0 \simeq a N^{\nu_3}$, where $a$ is the segment size and $N$ is the number of segments per chain. Unless otherwise specified, we are concerned with the good solvent condition in three-dimensional space and adopt for simplicity the mean-field value of the critical exponent $\nu_3 = 3/5$, where the swelling as against the ideal chain $\nu_3 > 1/2$ is due to the excluded-volume interactions among segments.
Increasing the volume fraction $\phi$ of the segment, one reaches the point 
$\phi^{(ov)} \simeq N^{-4/5}$ above which different polymers start to overlap in space.
In this so-called semidilute regime, the excluded-volume interactions are screened at the scale of the correlation length 
$\xi \simeq  a \phi^{-3/4}$, and the chain conformation can be envisioned as a random walk of the correlation blob of the size 
$\xi \simeq a g^{3/5}$ with $g$ segments inside, so that the overall size is given by 
$R_0 \simeq \xi ( N/g)^{1/2} \simeq a \phi^{-1/8} N^{1/2}$.
In the melt, which is a dense limit of the semidilute solution $\phi \rightarrow 1$, the excluded-volume interactions are screened at the segment scale, so the chain follows the almost ideal random walk statistics.

Turning our attention to the melt of ring polymers, the excluded-volume interactions would be screened as well, but for this very reason, the topological constraints, which may be buried under the excluded-volume effect in other situations, stand out. Therefore, unlike the linear polymer case, we encounter increased complexity instead of simplification due to the screening of the excluded-volume effect.
Indeed, it has been acknowledged in recent years that understanding conformations and dynamics of the melts of noncatenated ring polymers is one of the remaining challenges in the polymer science~\cite{Rubinstein_08,Vettorel,Halverson_2011}.
Towards this ambitious goal, the construction of a phenomenological theory would offer promising prospects, in which the topological constraints could be accounted through the geometrical picture, just as in the tube model and the reptation model in the theory of linear polymer melts.
The aim of the present paper is to provide a detailed account of the recently proposed such an attempt for the statistics of ring polymer melts~\cite{Sakaue_2011}.

The rest of the paper is organized as follows.
In Sec.~\ref{sec_mean_field}, we start with some preliminary remarks on the topological length scale, and then introduce a mean-field model. Sec.~\ref{Results} summarizes the results including (i) conformational properties of individual rings, (ii) global geometrical structure of melts and (iii) the effect of the chain stiffness along with the comparison with available numerical/experimental data. We shall also look into previous conjectures in light of the present theory. Finally, we close in Sec.~\ref{Conclusion} with summary and perspectives.

\section{Mean-field theory}
\label{sec_mean_field}
\subsection{Topological length scale}
\label{subsec_topo_length}
Entanglements formed in linear polymer melts and topological constraints in ring polymer melts differ in their nature, but they share some similarities which may trace back to the fact that both arise from the non-crossability constraint in flexible polymer systems.

It is well known that in melts of short linear polymers $N<N_e$, there are no signs of entanglement, and their dynamics is well described by the Rouse model~\cite{deGennes,Doi_Edwards,Rubinstein_Colby}. The entanglement length $N_e$ depends on the chain stiffness, concentration, etc., but its typical value in melts of flexible polymer is on the order of $N_e \sim 100$~\cite{Everaers_04}.
There is a quantity similar to $N_e$ for the ring polymer melts, which we call the {\it intrinsic topological length}  and denote by $N_1$ in the following. It has been observed in numerical simulations that in melts of short ring polymers $N<N_1$, there seems to be no sign of the topological constraints, and each ring obeys the ideal statistics with the size $R \simeq a N^{1/2}$~\cite{Vettorel,Halverson_2011,Muller2,Muller3,Suzuki}.
There are some good evidence that the numerical value of $N_1$ is comparable to that of $N_e$ in the corresponding linear chain melts~\cite{Vettorel,Halverson_2011}. We provide a geometrical interpretation of $N_1$ in Sec.~\ref{Large_Scale}.

In linear polymer melts with $N\gg N_e$, chains are mutually entangled, where the tube size is given by $\xi_{e} \simeq a N_e^{1/2}$. The tube size and, therefore, $N_e$ as well are independent of $N$ for $N \gg N_e$, which are important quantities in the rheological characterization of the system.
However, the analogy may break down for ring polymer melts with $N\gg N_1$ due to the non-trivial ring conformation in large scale in contrast to the ideal conformation in linear polymers. Since there is no reason {\it a priori} to claim the constant topological length,  we let it be $N$ dependent and denote it by $g_{\natural}(N)$ with the ``initial condition" $g_{\natural}(N_1) = N_1$.
The functional form of $g_{\natural} (N)$ and the conformation of individual rings at large scale $r > a [g_{\natural} (N)]^{1/2}$ for melts of high molecular weight rings $N > N_1$ should be determined from the topological constraints in the system. This can be done self-consistently by the following mean-field theory.

\subsection{Excluded-volume analogy and Order parameter}
\label{OrderParameter}
We are concerned with the topological effects in melts of noncatenated, unknotted rings. All other effects are irrelevant, which leads to the ideal behaviors in melts of linear polymers/phantom ring polymers. Two contributions could be identified. One is the inter-molecular (avoiding to link with other rings), and the other is the intra-molecular (avoiding the self-knotting). 
To write down the free energy for these effects, we recall the observation that two unlinked ring polymers with zero thickness, i.e., no excluded- volume, repel each other at close distances~\cite{Grosberg_Khokhlov,Frank-Kamenetskii}. The second virial coefficient of interaction between such rings is found to be of the order of the size of the ring cubed. That is to say, in dilute solutions, the topological effect restricting the phase space volume can be effectively treated as the excluded-volume problem. There is also a conjecture based on the analysis of Gauss linking number of random flexible rings that the topological constraints in long flexible rings and the excluded-volume interactions have similar effects~\cite{desCloizeaux}. We shall seek for a possible way to implement this excluded-volume analogy in the statistics of melts and concentrated solutions.

In what follows, we include the concentration effect by regarding a semidilute solution as a melt composed of the correlation blob of size $\xi \simeq  a \phi^{-3/4}$. The melt is understood to be the limit $\phi \rightarrow 1$.
The equilibrium spatial size $R$ (such as a radius of gyration) of each ring is a function of $N$ and the monomer volume fraction $\phi$.

Let us introduce several quantities relevant for the characterization of ring polymer melts.
The total number of monomers in the region of volume $\sim R^3$ of a single ring is $\sim R^3 \phi/a^3$, thus, the number of rings there is
\begin{eqnarray}
N_R \simeq \frac{R^3 \phi}{a^3 N} 
\label{N_x}
\end{eqnarray}
For a given ring, $N_R$ can roughly be thought of as the number of neighboring rings, so we call it the {\it coordination number}, which turns out to be an important measure for the geometrical characterization of the melts.
Equally important is the {\it self-density} 
\begin{eqnarray}
\phi_s \simeq \frac{a^3 N}{R^3}
\label{phi_s}
\end{eqnarray}
 which measures the degree of compactness of individual rings, or the degree of mutual ring interpenetration. These two quantities are related as $N_R \simeq \phi/\phi_s$.

The effective excluded-volume $v_R$ of the ring with the spatial size $R$ scales with volume and we write it as
\begin{eqnarray}
v_R\simeq R^3 Y
\label{v_R}
\end{eqnarray}
where a dimensionless factor $Y$ (independent of $N$) is introduced to account for the fact that the ring is not a rigid particle. Instead, it is a very soft object, and significant interpenetrations should occur in melts.
Of course, this factor $Y$ cannot be arbitrary, and shall be fixed later by the physical requirement associated with the intrinsic topological length $N_1$ mentioned in Sec.~\ref{subsec_topo_length}.
The ``volume fraction" of topological origin then follows as
\begin{eqnarray}
\phi_R \equiv \frac{v_R N_R}{R^3}= N_R Y
\label{phi_R}
\end{eqnarray}
 which plays a role as an order parameter in the theory.
We thus write down the free energy as a function of $\phi_R$ as
\begin{eqnarray}
F(\phi_R) = F_{inter}(\phi_R) + F_{intra}(\phi_R)
\label{free_energy_basis}
\end{eqnarray}

\subsection{Noncatenation constraint}
\label{Noncatenation_constraint}
Evaluating the noncatenation constraint contribution to the free energy is the most difficult part in the problem. 
We follow the excluded-volume analogy and make a simple proposal~\cite{Sakaue_2011}.
 Given an effective volume fraction $\phi_R$ of topological origin, we assume that its impact can be treated within the framework of liquid theory, where the excluded-volume effect in dense solutions has been intensively studied.
We adopt a classical van der Waals theory of fluids whose free energy density for one component fluid with the volume fraction $\phi$ and the excluded-volume $v$ is given by $f(\phi)/(k_BT) = [\phi/v] \ln{[\phi/(1-\phi)]} -\epsilon \phi^2$ with $k_B T$ being thermal energy.
In our case, this may be transformed as $f_{inter}(\phi_R)/(k_BT) \simeq -[\phi_R/v_R] \ln{[(1-\phi_R)]}$
where we set $\epsilon=0$ (athermal). The ideal gas term is irrelevant here, since it is already included in the non-topological part of the free energy.
The free energy per ring $F_{inter} =f_{inter} \times R^3/N_R$  is thus
\begin{eqnarray}
\frac{F_{inter}(\phi_R)}{k_B T } = - \ln{(1-\phi_R)}
\label{F_inter}
\end{eqnarray}

\subsection{Unknotting constraint}
\label{Unknotting_constraint}
 An inspection of eq.~(\ref{F_inter}) indicates that increasing the ring size $R$ is unfavorable due to the noncatenation constraint. 
Thus, this topology effect leads to the squeezing of the ring toward a globular state, which should be negotiated with the unknotting constraint. 
To derive the expression for $F_{intra}(\phi_R)$, it is instructive first to recall the free energy cost $\Delta F_{\Theta}$ to confine an isolated linear polymer into a small closed cavity under $ \Theta$ condition, where $\nu_3 = 1/2$. A standard scaling consideration suggests the following form;
\begin{eqnarray}
\frac{\Delta F_{\Theta}}{k_BT} \simeq \left( \frac{R_{\Theta}}{D}\right)^{\delta_{\Theta}}
\label{F_theta}
\end{eqnarray}
where $R_{\Theta} = a N^{1/2}$ is the unperturbed coil size and $D < R_{\Theta}$ is a linear dimension of the cavity.
The exponent $\delta_{\Theta}$ can be determined by examining the system size dependence of $\Delta F_{\Theta}$ under constant segment density (or constant osmotic pressure).
That is to say, we require $\Delta F_{\Theta} \rightarrow k \Delta F_{\Theta}$ upon enlarging the system size  as $N \rightarrow kN$ and $V \simeq D^3 \rightarrow k V$. This thermodynamic consideration fixes $\delta_{\Theta}=6$~\cite{Grosberg_Khokhlov,Sakaue_Raphael}.

We expect that the probability of self-knotting depends on the self-density $\phi_s$, and it linearly increases with the system size $N$ under constant $\phi_s$. Therefore, the above reasoning for $\Delta F_{\Theta}$ can be directly applied to $F_{intra}$, too, so we obtain
\begin{eqnarray}
\frac{F_{intra}}{k_BT} \simeq \left( \frac{R_{0}}{R}\right)^{6} \simeq \left( \frac{a}{R}\right)^6 N^3 \phi^{-3/4}
\label{F_intra_0}
\end{eqnarray}
where $R_0 \simeq  a \phi^{-1/8} N^{1/2}$ is the unperturbed polymer size in semidilute regime as already stated in Sec.~\ref{Introduction}, and $R<R_0$ is a linear dimension of individual rings squeezed by surrounding rings.
Using the definition of $\phi_R$ (eq.~(\ref{phi_R})) and $N_R$ (eq.~(\ref{N_x})), the above equation can be rewritten as
\begin{eqnarray}
\frac{F_{intra}(\phi_R);N}{k_BT} \simeq  \frac{N Y^2 \phi^{5/4}}{\phi_R^2}
\label{F_intra}
\end{eqnarray}
which guarantees the linear dependence on $N$ under the fixed order parameter $\phi_R$.

To get a physically appealing insight, let us again come back to the problem of the confined $\Theta$-chain. Its spatial structure can be viewed as a dense piling of blobs of size $\xi_{\Theta} \simeq a g_{\Theta}^{1/2}$ with $g_{\Theta}$ segments inside. This implies the relation $a^3g_{\Theta}/\xi_{\Theta}^3 \simeq \phi \simeq a^3 N /R^3  \Leftrightarrow \xi_{\Theta} \simeq a \phi^{-1}$ and $g_{\Theta}\simeq \phi^{-2}$. The free energy cost (eq.~(\ref{F_theta})) can be obtained by assigning $\sim k_BT$ per blob, i.e., $\Delta F_{\Theta}/k_BT \simeq N/g_{\Theta} \simeq R^3/\xi_{\Theta}^3$~\cite{Grosberg_Khokhlov,Sakaue_Raphael}.

It is tempting to follow the same line in our problem, too, but now the self-density $\phi_s$ should be used to write down the space-filling condition. Let us do so. A view of the compact ring of size $R \ (<R_0)$ as a dense piling of topological mesh (blob) implies $\xi_{\natural} \simeq \xi (g_{\natural}/g)^{1/2}$ and $a^3 g_{\natural}/\xi_{\natural}^3 = \phi_s$. One thus finds
\begin{eqnarray}
&& \xi_{\natural} \simeq a \phi^{1/4} \phi_s^{-1} \simeq a \phi^{-3/4} N_R \simeq \xi N_R \label{xi_topo}\\
&& g_{\natural} \simeq \phi^{3/4}\phi_s^{-2} \simeq \phi^{-5/4} N_R^2 \simeq g N_R^2. \label{g_topo}
\end{eqnarray}
Assigning $\sim k_BT$ per topological strand, we obtain $F_{intra}/(k_BT) \simeq N/g_{\natural} \simeq  R^3/\xi_{\natural}^3$, which coincides with the scaling derivation of $F_{intra}$ (eq.~(\ref{F_intra})). 
Moreover, the deduced relations $\xi_{\natural} \simeq \xi N_R$ and $g_{\natural} \simeq g N_R^2$ point to an interesting relation between the topological constraint and the geometrical property of the melt. Indeed, it indicates that the number of units (correlation blobs) required to form a topological constraint is of order $\sim N_R^2$.


\subsection{Short ring case and Y-factor}
\label{Y-factor}
To complete the formulation, the factor $Y$ introduced in eq.~(\ref{v_R}) remains to be fixed.
This can be done by considering the behavior of the short ring melts with $N < N_1$ (or the short scale behaviors of longer ring melts).

To begin with, let us recall the logical structure of the mean-field theory.
The free energy given in eq.~(\ref{free_energy_basis}) is a so-called constraint free energy obtained by partial summation over phase space with $\phi_R$ (order parameter) fixed. The true thermodynamic free energy ${\mathcal F}$ is given by 
\begin{eqnarray}
\exp{[- \beta {\mathcal F}] } = \int d\phi_R \ \exp{[-\beta F(\phi_R)]} \nonumber
\end{eqnarray}
where $\beta = (k_B T)^{-1}$.
For long ring solutions ($N \gg N_1$), i.e., {\it large systems}, the above integral is dominated by the narrow region around the minimum $\phi_R = \phi_R^{(eq)}$, leading to
\begin{eqnarray}
{\mathcal F} \simeq F(\phi_R^{(eq)}) \nonumber
\end{eqnarray}
and $\phi_R^{(eq)}$ represents the equilibrium value of $\phi_R$ from which $N_R^{(eq)} = \phi_R^{(eq)} /Y$ and $R^{(eq)} \simeq a N^{1/3} (N_R^{(eq)})^{1/3} \phi^{-1/3}$ follow; see eq.~(\ref{N_x}) and~(\ref{phi_R}). Note that superscript $(eq)$ marking the equilibrium value will be attached only when it is necessary, and will be omitted unless confusion is expected. 
However, for the melts with short rings, i.e., {\it small systems}, fluctuations around the minimum are substantial, and the above minimization procedure is insufficient to capture the equilibrium properties of the system. 

\begin{figure}[h]
\includegraphics[width=0.45\textwidth]{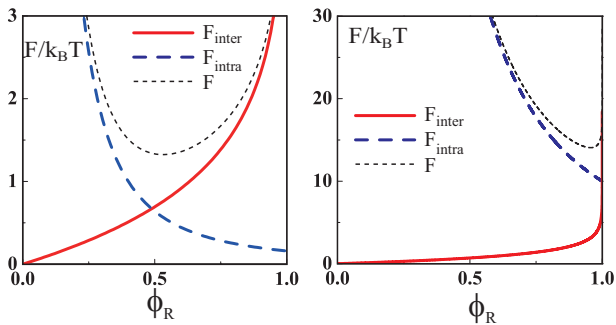}
\caption{(Color online) Free energy profiles as a function of $\phi_R$ for the melt with short rings $N/g^*=(0.4)^2$ (left) and long rings $ N/g^*=10$ (right). }
\label{Fig1}
\end{figure}

For example, we plot in Fig.~\ref{Fig1} (left) the free energy profiles for the melt of short rings ($N \simeq N_1$), where the free energy minimum is attained at low $\phi_R < 1$, and the corresponding free energy is comparable to the thermal energy. Physically, this means that for melts with shorter rings ($N < N_1$), the topological constrains are buried in the thermal fluctuations, therefore, ideal chain behavior is expected there.
One may find a situation $F_{inter}({\tilde \phi}_R) \simeq F_{intra}({\tilde \phi}_R; \ N_{1}) \simeq k_BT$ at ${\tilde \phi}_R$.
From this, we find the expression for the factor $Y$;
\begin{eqnarray}
Y \simeq N_{1}^{-1/2} \phi^{-5/8} \  {\tilde \phi}_R \simeq \left( \frac{N_1}{g}\right)^{-1/2} {\tilde \phi}_R \simeq \left(\frac{\xi_1}{\xi} \right)^{-1} {\tilde \phi}_R
\label{eq:Y}
\end{eqnarray}
where ${\tilde \phi}_R$ is an onset value of the order parameter above which the topological constraints set in. The value of ${\tilde \phi}_R$, which must be smaller than unity, is a universal number in our theory in the sense that its precise value is insensitive to the structural details of individual polymer systems, rather determined by a geometrical requirement (see Sec.~\ref{Large_Scale}). Figure~\ref{Fig1} (left) indicates a reasonable range for it around ${\tilde \phi}_R \sim 0.5$. Later, in Sec.~\ref{Large_Scale}, we shall see that ${\tilde \phi}_R \simeq 0.4$ can indeed account for recent numerical simulations very well. In the last expression,  $\xi_1 \simeq \xi (N_1/g)^{1/2} \simeq a N_1^{1/2}\phi^{-1/8}$ is the spatial length scale corresponding to $N_1$. 

As expected, the $Y$-factor depends on the intrinsic topological length $N_1$ and acts as a {\it softening} factor compared to the rigid particle, which allows the partial interpenetration of different rings.
By lumping $N_1$ and ${\tilde \phi}_R$ together, one can rewrite eq.~(\ref{eq:Y}) as
\begin{eqnarray}
Y \simeq \left( \frac{g^*}{g}\right)^{-1/2} \simeq \left( \frac{\xi^*}{\xi}\right)^{-1}
\label{eq:Y-2}
\end{eqnarray}
where 
\begin{eqnarray}
\xi^* \equiv \frac{\xi_1}{{\tilde \phi}_R} \simeq \xi \left( \frac{g^*}{g}\right)^{1/2}, \quad g^* \equiv \frac{N_{1}}{{\tilde \phi_R}^2}
\label{g*}
\end{eqnarray}
 are the size of the topological mesh and the number of segments in it, respectively, in the compact statistic regime as detailed below in Sec.~\ref{Results}.

The intra-molecular free energy eq.~(\ref{F_intra}) can now be rewritten as
\begin{eqnarray}
\frac{F_{intra}(\phi_R; N)}{k_BT} = \left(\frac{\phi_R}{{\tilde \phi_R}} \right)^{-2} \frac{N}{N_{1}} = \phi_R^{-2} \frac{N}{g^*}
\label{F_intra_2}
\end{eqnarray} 

\section{Results and Discussion}
\label{Results}
Now the stage is set, so we proceed to seek for the spatial structure of rings in large scales $N > N_{1}$. In the following, whenever concrete numerical calculations are necessary, we set all numerical prefactors of order unity (indicated by the $\simeq$ symbols in equations) to be unity.
As shown in Fig.~\ref{Fig1} (right), here, the energy scale is much larger than the thermal energy, thus the equilibrium size can readily be determined by minimizing the free energy with respect to $\phi_R$. 

\subsection{Conformation of individual rings}
\label{conformation}
\begin{figure}[h]
\includegraphics[width=0.37\textwidth]{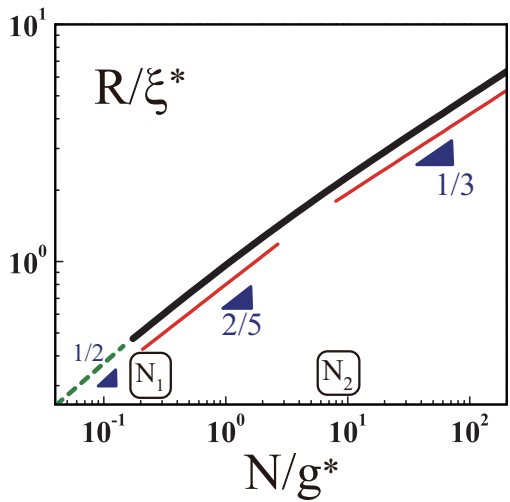}
\caption{(Color online) Master curve of the ring size $R/\xi^*$ as a function of the chain length $N/g^*$ (double logarithmic scale) obtained by numerically minimizing the free energy eq.~(\ref{free_energy_basis}). }
\label{Fig2}
\end{figure}
In Fig.~\ref{Fig2}, we plot the equilibrium size $R$ of rings as a function of the polymerization index $N$. The following trends can be read off; (i) sufficiently long rings ($N>N_2$) obey compact statistics with $\nu=1/3$, (ii) rings with intermediate length ($N_{1}<N<N_2$) are well characterized by the effective exponent $\nu_{eff} \simeq 2/5$.
Therefore, one can say that the statistics of ring polymer melts have two well-defined regimes: ideal and compact statistics for short $N<N_{1}$ and long $N>N_2$ rings, respectively, which are separated by a wide crossover region spanning over more than one order of magnitude.

To get an insight into this crossover, we note that from eqs.~(\ref{N_x}),~(\ref{phi_R}),~(\ref{F_inter}),~(\ref{eq:Y}),(\ref{F_intra_2}), a mean-field solution for $N>N_{1}$ can generally be written in the form
\begin{eqnarray}
R= \xi^* \left(\frac{N \phi_R^{(eq)}}{g^*}\right)^{1/3} 
\label{R_formula_1}
\end{eqnarray}
where $\xi^* \equiv a \phi^{-1/8} g^{*1/2}  = \xi_1 {\tilde \phi}_R^{-1}$ (eq.~(\ref{g*})), and $\phi_R^{(eq)}$ denotes the equilibrium value of the order parameter which depends solely on $N/g^*$, i.e., $\phi_R^{(eq)} = \phi_R^{(eq)}(N/g^*)$.
The above equation can be transcribed into a physically appearing form as
\begin{eqnarray}
R= \xi_{\natural}(N) \left(\frac{N}{g_{\natural}(N)}\right)^{1/3} 
\label{R_formula_2}
\end{eqnarray}
where the subtle dependence of $\phi_R^{(eq)}$ on $N$ is absorbed in the topological length as $\xi_{\natural}(N) = \xi^* \phi_R^{(eq)}$ and $g_{\natural}(N) = g^* (\phi_R^{(eq)})^2$.
This indicates that individual rings are squeezed by the noncatenation constraints, and hence, follow the compact statistics {\it at fixed} $N$. In addition, unlike the usual globule found in liner polymers, the unknotting constraint of rings leads to a unique conformation known as a {\it crumpled globule}~\cite{crumpled_globule}, i.e., upon renormalization a large scale spatial organization of individual rings is formally characterized by a space-filling curve with fractal dimension $D_f = \nu^{-1} = 3$ (see Sec.~\ref{comparison}).
As this expression shows, the peculiarity in the ring conformation in melts lies in the fact that a renormalization factor, which may be called {\it topological mesh}~\cite{crumpled_globule}, $\xi_{\natural}(N) = a g_{\natural}^{1/2}(N) \phi^{-1/8}$ is {\it length dependent} as a result of the competition between inter-molecular and intra-molecular topological constraints.

\begin{figure}[h]
\includegraphics[width=0.45\textwidth]{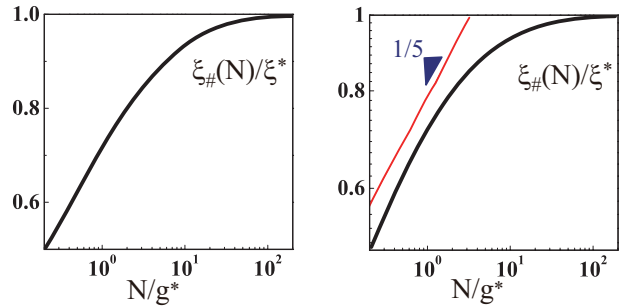}
\caption{(Color online) Topological mesh size $\xi_{\natural}/\xi^*$ as a function of the chain length $N/g^*$.  (Left) is a semi-log plot showing a logarithmic increase towards to $\xi_{\natural} \rightarrow \xi^*$, while (Right) is a double logarithmic plot showing an effective power low $\xi_{\natural} \sim N^{1/5}$ in the intermediate range of the chain length ($ N_1 < N < N_2$).  }
\label{Fig_xi_N}
\end{figure}

Looking at the behavior of the order parameter, it takes the value $\phi_R^{(eq)} = {\tilde \phi}_R$ at the onset of the topological constraints ($N=N_1$), and increases with $N$ towards the saturation $\phi_R^{(eq)} \rightarrow 1$ at $N \rightarrow N_2$, where the saturation point $N=N_2$ can be identified as the onset of the compact statistics. Correspondingly, $g_{\natural}$ and $\xi_{\natural}$ change in such a way that $ g_{\natural}=N_1$ and $\xi_{\natural} = \xi_1$ at $N=N_1$ and they increase with $N$ towards the saturation  $ g_{\natural}=g^*$ and $\xi_{\natural} = \xi^*$ at $N=N_2$.
As $N_2 \gg g^* = N_1 {\tilde \phi}_R^{-2}$ (a formula to estimate $N_2$ will be given in eqs.~(\ref{N_2_1}),~(\ref{N_2_p_1}) and~(\ref{N_2_p_2}) along with its physical interpretation), the evolution of $g_{\natural}$, $\xi_{\natural}$ during this interval is quite slow. Indeed, this slow evolution is controlled by the logarithmic divergence towards a topological dense-packed limit ($\phi_R \rightarrow 1$), but well fitted by the effective (fictive) power laws $\xi_{\natural} \sim N^{1/5}$ and $g_{\natural} \sim N ^{2/5}$ (Fig.~\ref{Fig_xi_N}).
This renders a small correction to the exponent $\nu$, which leads to the effective exponent in the ring size $R \simeq \xi_{\natural} (N/g_{\natural})^{1/3} \sim N^{2/5}$. Such a feature has been observed in several numerical simulations~\cite{Muller2,Muller3,Pakula,Muller1,Brown} and real experiments~\cite{Arrighi} with moderate chain length, which indeed coincides with the Cates-Deutsch conjecture~\cite{C-D}. However, we emphasize that $\nu_{eff} \simeq 2/5$ found here is not a well-defined exponent in our theory, but a fictive exponent seen in a broad (but restricted) range of the chain length $N_1 < N < N_2$.

For large enough molecular weight ($N > N_2$), the compact statistics $R= \xi^* (N/g^*)^{1/3}$ is realized in accordance with recent numerical and experimental observations~\cite{Vettorel,Halverson_2011,Suzuki,Takano}. The coordination number follows from eq.~(\ref{N_x}) as $N_R \simeq g^{*1/2} \phi^{5/8}$, and thus the self-density $\phi_s \simeq \phi/N_R \simeq g^{*-1/2} \phi^{3/8}$ (eq.~(\ref{phi_s})). This agrees with results for the concentrated solution of randomly branched polymers derived by Daoud and Joanny~\cite{Daoud_Joanny,Khokhlov_Nechaev}. Their expressions for the chain size, the coordination number and the self-density in melt limit $\phi \rightarrow 1$ are, respectively, $R \simeq a N^{1/d} \Lambda^{(2-d)/2d}$, $N_R \simeq \Lambda^{(2-d)/2}$ and $\phi_s \simeq \Lambda^{(d-2)/2}$  for the space dimension $2 < d < 4$, where $\Lambda^2$ is called the activity of the trifunctional units (branching points). Noting that $\Lambda^{-1} \simeq g^*$ corresponds to the linear parts between trifunctional units, our results are recovered by inserting $d=3$. It is interesting to observe that $N_R$ (the degree of the overlap) is controlled by $\Lambda$ (the degree of the branching). Indeed, $N_R$ increases with decreasing $\Lambda$, and crosses over to the value of linear chain melts $N_R \simeq N^{(d/2)-1}$  at $\Lambda \simeq N^{-1}$, where a labeled chain is surrounded by a large number of other chains, which grows as $N^{1/2}$ in $d=3$, leading to significant entanglements.

The optimum free energy in this compact statistic regime is
$F/(k_B T) \simeq N/g_{\natural}$
, i.e., an order of the magnitude of thermal energy per topological strand.
This translates into the osmotic pressure
$\Pi_{top} \simeq F N_R/R^3 \simeq k_B T \phi/(a^3 g_{\natural}) \simeq (N_R)^{-2} k_BT \phi^{9/4}/a^3$.
Adding this to the usual excluded-volume contribution $\Pi_{0} \simeq k_BT/\xi^3 \simeq k_BT \phi^{9/4}/a^3$~\cite{deGennes}, the total osmotic pressure thus can be written as
\begin{eqnarray}
\Pi = \Pi_0 + \Pi_{top} \simeq (1+N_R^{-2}) \Pi_{0} .
\end{eqnarray}
It would be instructive to compare this result with the change in the osmotic pressure in linear chain solutions due to an increment in the local excluded-volume.
Let $v \simeq a^3$ and $c$ be the excluded-volume and the concentration, respectively, so that $\phi = c v$. Then, the osmotic pressure is $\Pi_0/k_BT   \simeq  v^{-1} \phi^{9/4} = v^{5/4}c^{9/4}$. Let us slightly increase the excluded-volume as $v \rightarrow v+ \delta v$ ($\delta v/v \ll 1$). The resultant increase in osmotic pressure is $\delta \Pi_0(v) = \Pi_0(v+\delta v)-\Pi_0(v) \simeq \Pi_0(v) \delta v/v$.
Therefore, the effect of topological constraints seems to be reflected as an increase of the local excluded-volume interactions, i.e.,  $N_R^{-2} \leftrightarrow \delta v/v$ in accordance with the des Cloizeaux conjecture~\cite{desCloizeaux}.
By inserting the value of the coordination number estimated below (Sec.~\ref{Large_Scale}), we see that the additive term is too small to be detected in practical measurements~\cite{Halverson_2011}.

\subsection{Large-scale spatial structure}
\label{Large_Scale}
\begin{figure}[h]
\includegraphics[width=0.38\textwidth]{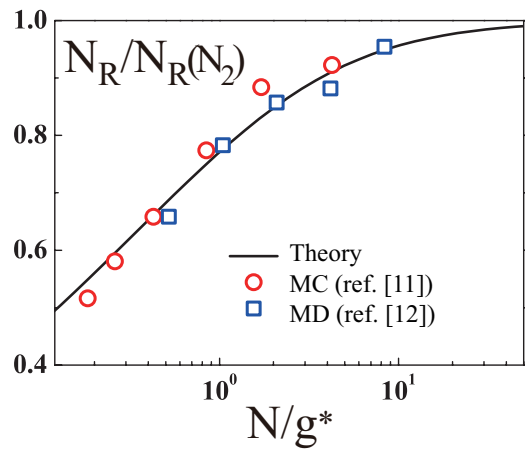}
\caption{(Color online) Dependence of the coordination number $N_R$ on the chain length $N$. The solid line is the theoretical prediction, while points are results from numerical simulations (read off from Fig. 8 in ref.~\cite{Vettorel}. and Fig. 13 in ref.~\cite{Halverson_2011}). Circle: Monte Carlo simulation by Vettorel et. al. ($N_e \simeq 175,\ N_R(N_2) \simeq 15$)~\cite{Vettorel}, Square: Molecular dynamic simulation by Halverson et. al. ($N_e \simeq 28, \ N_R(N_2) \simeq 17$)~\cite{Halverson_2011}.}
\label{Fig_NR_compare}
\end{figure}

During the broad crossover region ($N_1 < N < N_2$), the melt of rings adjust its own spatial structure by balancing the noncatenation and unknotting topological constraints. The coordination number $N_R$ (eq.~(\ref{N_x})) is a good measure to capture the large scale ($r>R$) geometrical feature of the melts. 
According to eqs.~(\ref{phi_R}) and~(\ref{eq:Y}), it is expressed as 
\begin{eqnarray}
N_R(N) = \frac{\phi_R(N)}{Y} \simeq \left( \frac{N_1}{g} \right)^{1/2}  \frac{\phi_R(N)}{{\tilde \phi}_R}
\label{N_R-phi_R}
\end{eqnarray}
so that it slowly increases from $N_R (N_1)$ at $N=N_1$ towards $N_R (N_2)$ at $N=N_2$.
These two values $N_R(N_1)$ and $N_R(N_2)$ turn out to be key quantities in the following discussion.
\begin{itemize}
\item At the onset of topological constraints ($N=N_1$): We have $\phi_R(N_1) = {\tilde \phi}_R$ so that
\begin{eqnarray}
N_R(N_1) = \frac{{\tilde \phi}_R}{Y} \simeq \left( \frac{N_1}{g}\right)^{1/2} \simeq N_1^{1/2}\phi^{5/8}
\label{N_R_N_1_item}
\end{eqnarray}
Hence, $N_R(N_1)$ is found to be directly linked to the intrinsic topological length $N_1$. A rough estimate $N_1 \sim 100$, $\phi \sim 0.5$ gives $N_R(N_1) \sim 6$.

\item At the onset of compact statistics ($N=N_2$): We have $\phi_R(N_2) = 1$ so that
\begin{eqnarray}
N_R(N_2) = \frac{1}{Y} = \frac{N_R(N_1)}{{\tilde \phi}_R} \simeq \left( \frac{g^*}{g}\right)^{1/2} \simeq g^{*1/2}\phi^{5/8}
\label{N_R_N_2_item}
\end{eqnarray}
A rough estimate $N_R(N_1) \sim 6$, ${\tilde \phi}_R \sim 0.4$ gives $N_R(N_2) \sim 15$, which is in reasonable agreement with recent numerical simulations (see the caption of Fig.~\ref{Fig_NR_compare}).
Its geometrical interpretation was suggested in ref.~\cite{Vettorel}, where the correspondence of $N_R(N_2)$ with the coordination number of jammed hard ellipsoid systems~\cite{Chaikin_06} was pointed out.
Hence, $N_R(N_2)$ is supposed to be linked to the geometrical property inherent in the packing in three-dimensional space.

\end{itemize}

According to eq.~(\ref{N_R_N_2_item}), the above consideration identifies the Y-factor $\simeq 1/N_R(N_2)$ to be a geometrical quantity, too, which in turn suggests the geometrical origin of the intrinsic topological length scale $N_1$ (eq.~(\ref{eq:Y})), thus also $N_R(N_1)$ as well as $g^*$ (eq.~(\ref{eq:Y-2})).

In Fig.~(\ref{Fig_NR_compare}), we compare the prediction of our mean-field theory with the results from recent numerical simulations. To determine the scaling factor $g^*$, we first identify the intrinsic topological length $N_1$ to be equivalent to the entanglement length $N_e$ which was measured in simulations, and then assigned ${\tilde \phi_R} = 0.38$ for the order parameter at which the topological constraints set in. This yields $g^* \simeq 1200$ for the model in ref.~\cite{Vettorel} and $g^* \simeq 190$ for that in ref.~\cite{Halverson_2011}. It is remarkable that with the above procedure for the rescaling, i.e., (i) the identification of $N_1 = N_e$, (ii) the assignment of a reasonable and common value of ${\tilde \phi}_R$, two different sets of simulation data are almost perfectly collapsed into the theoretical master curve. The fact that the common ${\tilde \phi}_R$ value can fit the results from different simulation models (flexible rings in lattice Monte Carlo simulation~\cite{Vettorel} and semi-flexible rings in off-lattice Molecular dynamics simulation~\cite{Halverson_2011}) indicates that ${\tilde \phi}_R$ is insensitive to the model details. This is an expected result from the following consideration; From eq.~(\ref{phi_R}) with $\phi_R^{(eq)}(N_1)= {\tilde \phi}_R$ and $\phi_R^{(eq)}(N_2)=1$, it follows that
\begin{eqnarray}
{\tilde \phi}_R = \frac{N_R(N_1)}{N_R(N_2)}
\label{tilde_phi_N_R}
\end{eqnarray}
As already described above, $N_R(N_1)$ and $N_R(N_2)$ are geometrical quantities inherent to three-dimensional space, so their ratio should be universal, too.

\subsection{Physical picture of the crossover}
\label{physical_picture}
\begin{figure}[h]
\includegraphics[width=0.43\textwidth]{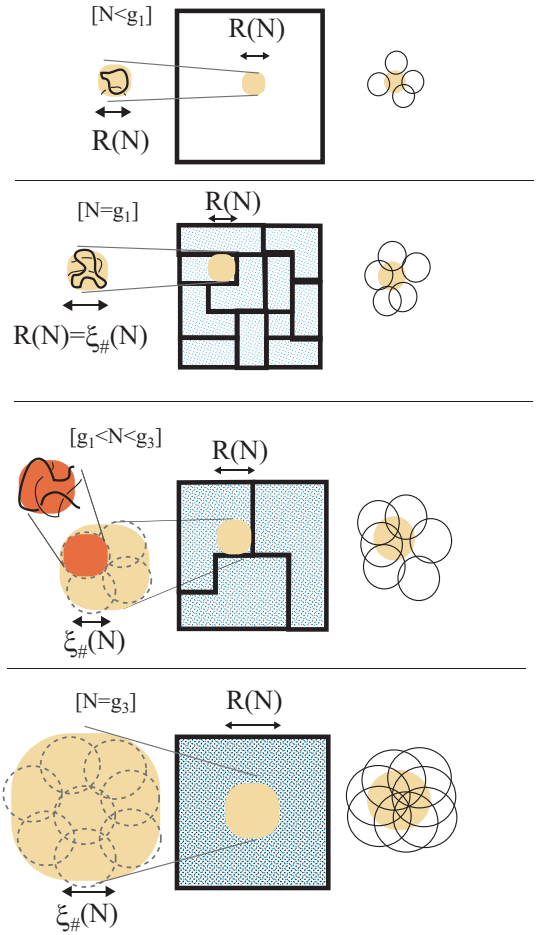}
\caption{(Color online) Schematics illustrating the spatial structures in ring polymer melts. [top] almost ideal ring regime ($N<N_1$), [2nd line] onset of the topological constraints ($N=N_1$), [3rd line] crossover regime ($N_1 < N < N_2$), [bottom] onset of the compact statistics ($N=N_2$).  The right column shows a reference ring (shaded) and its neighbors, which are represented by soft spheres. During the crossover, the coordination number $N_R$, thus, the degree of the overlap measured by $\phi_R$ increases with $N$ as shown in the right column. 
The middle column represents the global structure focusing on how the size and the correlation of topologically active clusters grow with $N$.
The left column shows the internal structure (meshes represented by dotted circles) of individual rings, and the finer inter-mesh structure with (a part of) the reference ring (thick line) and a part of other neighboring ring (thin line) penetrating the reference ring region.}
\label{Fig_schematics}
\end{figure}

After clarifying both the intra- and inter-molecular spatial structures, we now ask what is the underlying physical picture of the broad crossover regime ($N_1 < N < N_2$).
Let us look at one given ring (shaded circle in Fig.~\ref{Fig_schematics}) in the melt and call it a reference ring. Define the number 
\begin{eqnarray}
m(N) \equiv \frac{N}{g_{\natural}(N)}
\label{m_N}
\end{eqnarray}
 of the topological meshes per ring. It takes $m(N_1) = N_1/g_{\natural}(N_1) = 1$ at the onset of the topological effect $N=N_1$, and slowly increases to $m(N_2) = N_2/g_{\natural}(N_2)=N_2/g^*$ at $N=N_2$. At even larger $N > N_2$, $g_{\natural} (N)$ saturates, so that $m(N)$ linearly increase with $N$ in the compact statistic regime.  
These meshes are forced to form due to the intra-molecular topological repulsion against the squeezing by surrounding rings. Therefore, any topological mesh in the reference ring is under the interference of the surrounding rings, as schematically illustrated in the left column in Fig.~\ref{Fig_schematics}.

Let us further define the occupation rate 
\begin{eqnarray}
M(N) \equiv \frac{m(N)}{N_R(N)} \simeq \frac{N}{g [N_R(N)]^3} .
\label{M_N}
\end{eqnarray}
where the last relation follows from eqs.~(\ref{g_topo}) and~(\ref{m_N}).
It represents the fraction of the $N_R$ surrounding rings which take part in the interference with the reference ring, thus forming an active pair with the reference ring in the sense of the topological constraint.
At $N=N_1$, we have $M(N_1) \simeq 1/N_R(N_1) \ll 1$. This means that at the onset of the topological effect, there is some freedom to form a nascent mesh, i.e., it can be formed cooperatively, where all $N_R(N_1)$ rings moderately contribute to the net interference energy $\sim k_BT$, or alternatively it can be formed by an intense interaction (on the order of $\sim k_BT$) with any one of the $N_R(N_1)$ rings. In this situation, there is some combinatorial freedom to create the mesh.

Now suppose the hypothetical process in which we increase the chain length $N$ with $\phi_R$ fixed. It is clear from the free energy eqs.~(\ref{F_inter}),~(\ref{F_intra_2}) with their behaviors shown in Fig.\ref{Fig1} that the created state is unstable with high intra-molecular energy cost $F_{intra}$ with lots of meshes. Switching off the constant $\phi_R$ constraint, therefore, the system relaxes to the equilibrium state with larger $\phi_R$ by enlarging the ring size $R$ at the expense of the inter-molecular energy cost $F_{inter}$ with the larger coordination number $N_R$. The size of meshes $\xi_{\natural}$ also grows with $N$ so that the self-density $\phi_s \simeq a^3 g_{\natural}/\xi_{\natural}^3 \simeq (a/\xi_{\natural}) \phi^{1/4}$ decreases. This dilution of the self-density is caused by the interpenetration of the surrounding rings, i.e., the increase of $N_R$. In essence, the topological constraints become tighter with increasing $N$ as a result of the self-consistent spatial structural adjustment.

In fact, by comparing the growth rate of $m(N)$ and $N_R(N)$ (adopting fictive exponents obtained in Sec.~\ref{conformation}, these quantities scale as $m(N) \sim N^{3/5}$ and $N_R(N) \sim N^{1/5}$ in the crossover regime), one finds that  the occupation rate $M(N) \simeq N/(g [N_R(N)]^3)$ grows with $N$, and eventually reaches the saturation point, i.e., the onset of the compact statistic, which can be defined by the relation $M(N_2) \simeq 1$. From this condition, we find 
\begin{eqnarray}
 N_2 \simeq g \times [N_R(N_2)]^3 \simeq N_1 \times \left(\frac{N_R(N_2)^3}{N_R(N_1)^2} \right) 
 \label{N_2_1}
 \end{eqnarray}
 where the last relation follows from eq.~(\ref{N_R_N_1_item}).
 Substituting the estimates
$N_R(N_1) \sim 6$, $N_R(N_2) \sim 15$ and $N_1 = N_e$ from the simulation data, we find the estimates $N_2 \sim 10^4$ for the model in ref.~\cite{Vettorel} and $N_2 \sim 10^3$ for that in ref.~\cite{Halverson_2011}, in rather good agreement with observations.
 
Physically, this saturation condition means that there are $N_R$ meshes in the reference ring, all of which interfere with one of the surrounding $N_R$ rings. There is, on average, one-to-one correspondence between mutually interfering meshes. Hence, the crossover regime is characterized by the continuous loss of the combinatorial freedom to form the network of meshes, and $M(N) \simeq 1$ at $N=N_2$ signifies the densely packed (no voids) state in terms of the topological volume fraction $\phi_R \rightarrow 1$.

Figure~\ref{Fig_schematics} summarizes this subsection. At $N>N_1$, the system of ring polymer melts can be viewed as composing of topologically active clusters.
Rings belonging to the same cluster cooperatively create meshes within these rings. If we look at one ring (reference ring marked in the middle column) in a particular cluster, some of its neighboring rings belong to the same cluster, but others do not. The increase of the occupation rate $M(N)$, which is the fraction of the former, with $N$ corresponds to the growth of the cluster size. At $M(N=N_2) \simeq 1$, the combinatorial freedom for the clustering patterns is lost, and the whole system is spanned by a single cluster.

\subsection{Effect of stiffness}
\label{stiffness}
In linear chain melts, the chain stiffness is known to be an important factor controlling the degree of entanglements, i.e., a slight increase in the stiffness promotes it  by reducing $N_e$ substantially~\cite{Everaers_04}.
Similarly, in the ring polymer melts, it is known that tuning the persistence length provides an efficient route to increase the ring overlap~\cite{Muller2}. We shall examine such a stiffness effect here.

With the increase in the local chain stiffness, each statistically independent segment becomes anisotropic with the thickness $a$ and the length $l > a$. We define the polymerization index as $N = L/a$, where $L$ is the contour length of the ring. The number of statistical (Kuhn) segments per ring is thus $N_K = L/l=N/p$, where $p = l/a$ is the stiffness parameter (aspect ratio of the segment).
According to the standard theory of polymer solutions, there are two qualitatively different regimes which are separated at the threshold stiffness parameter $p^* \simeq \phi^{-1/3}$~\cite{Grosberg_Khokhlov,Schaefer,Sakaue_2007}. 
At $p <p^*$, the correlation along the chain can be treated through the blob as in the flexible chain solutions, which has now $p$ dependence as 
\begin{eqnarray}
\xi \simeq a \phi^{-3/4}p^{-1/4}, \quad g \simeq \phi^{-5/4}p^{-3/4} \qquad (p<p^*).
\label{xi_g_p}
\end{eqnarray}
 The unperturbed (i.e., no topological constraint) polymer size which appears in eq.~(\ref{F_intra_0}) is thus $R_0 \simeq \xi(N/g)^{1/2} \simeq a\phi^{-1/8}N^{1/2}p^{1/8}$.  On the other hand, at $p > p^*$, the correlation along the chain is negligible and the mean-field theory is reliable there. In this mean-filed regime, the chains without topological constraint exhibit nearly ideal statistics on all length scales larger than the segment length $l$, where it is convenient to define the quantities 
\begin{eqnarray}
 \Xi \simeq a \phi^{-1} p^{-1}= a G^{1/2}p^{1/2}, \quad G \simeq \phi^{-2}p^{-3} \qquad (p>p^*), 
\label{Xi_G_p}
\end{eqnarray} 
so that $R_0 \simeq \Xi(N/G)^{1/2} \simeq aN^{1/2}p^{1/2}$. In subsequent formulas below, these quantities play analogous roles as $g$ and $\xi$, respectively, in this mean-field regime $p > p^*$.

Following the construction of free energy in Sec.~\ref{Noncatenation_constraint} and~\ref{Unknotting_constraint}, it is easy to see that eq.~(\ref{F_inter}) for $F_{inter}$ remains intact, but the appearance of eq.~(\ref{F_intra}) for $F_{intra}$ is modified due to the $p$ dependence in $R_0$;
\begin{eqnarray}
\frac{F_{intra}(\phi_R);N}{k_BT} \simeq \left\{
           \begin{array}{ll}
              \frac{N Y^2 \phi^{5/4}p^{3/4}}{\phi_R^2} &  \quad ( p < p^*) \\
             \frac{N Y^2 \phi^{2}p^{3}}{\phi_R^2} &  \quad ( p > p^*)
           \end{array}
        \right. 
        \label{F_intra_p_0}
\end{eqnarray}
By following the argument in Sec.~\ref{Y-factor}, we can fix the Y-factor from the condition $F_{inter}({\tilde \phi}_R) \simeq F_{intra}({\tilde \phi}_R; \ N_{1}) \simeq k_BT$;
\begin{eqnarray}
Y \simeq \left\{
           \begin{array}{ll}
              {\tilde \phi}_R N_1^{-1/2} \phi^{-5/8} p^{-3/8} = \left(\frac{N_1}{g}\right)^{-1/2} {\tilde \phi}_R &   ( p < p^*) \\
             {\tilde \phi}_R N_1^{-1/2} \phi^{-1} p^{-3/2} = \left(\frac{N_1}{G}\right)^{-1/2} {\tilde \phi}_R &   ( p > p^*)
           \end{array}
        \right. 
\end{eqnarray}
which can be rewritten in the form of eq.~(\ref{eq:Y-2}) as
\begin{eqnarray}
Y \simeq \left(\frac{\xi^*}{\xi} \right)^{-1} \simeq \left\{
           \begin{array}{ll}
              \left( \frac{g^*}{g}\right)^{-1/2} &  \quad ( p < p^*) \\
             \left( \frac{g^*}{G}\right)^{-1/2} &  \quad ( p > p^*)
           \end{array}
        \right. 
\end{eqnarray}
with
\begin{eqnarray}
\xi^* \equiv  \frac{\xi_1}{{\tilde \phi}_R} \simeq \left\{
           \begin{array}{ll}
              \xi \left( \frac{g^*}{g}\right)^{1/2} &  \quad ( p < p^*) \\
             \xi \left( \frac{g^*}{G}\right)^{1/2} &  \quad ( p > p^*)
           \end{array}
        \right. 
\end{eqnarray}
Substituting this result into eq.~(\ref{F_intra_p_0}), it turns out that $F_{intra}$ can be generally written in the form of eq.~(\ref{F_intra_2}). Therefore, an important $p$ dependence can be absorbed in the intrinsic topological length $N_1$ (or equivalently $g^*$). 
Hence, all the results (including Figs~\ref{Fig2},~\ref{Fig_xi_N},~\ref{Fig_NR_compare}) expressed in the rescaled forms already encompass the stiffness effect.

For moderately long ring systems ($N>N_1$), the topological meshes are created. From the space filling condition of these $a^3 g_{\natural}/\xi_{\natural}^3 \simeq \phi_s$ (Sec.~\ref{Unknotting_constraint}) with the intra-mesh statistics $\xi_{\natural}\simeq \xi (g_{\natural}/g)^{1/2} \simeq a \phi^{-1/8}g_{\natural}^{1/2}p^{1/8}$ for $p<p^*$ and $\xi_{\natural}\simeq \Xi (g_{\natural}/G)^{1/2} \simeq a g_{\natural}^{1/2} p^{1/2}$ for $p>p^*$, we obtain the mesh size $g_{\natural}$ and $\xi_{\natural}$ in terms of the coordination number
 as follows;
\begin{eqnarray}
\xi_{\natural} \simeq \left\{
           \begin{array}{ll}
              \xi N_R &  \quad ( p < p^*) \\
             \Xi N_R &  \quad ( p > p^*)
           \end{array}
        \right. 
\end{eqnarray}
\begin{eqnarray}
g_{\natural} \simeq \left\{
           \begin{array}{ll}
              g N_R^2 &  \quad  ( p < p^*) \\
             G N_R^2 &  \quad  ( p > p^*)
           \end{array}
        \right. 
        \label{g_N_R_p}
\end{eqnarray}
which are generalizations of eqs.~(\ref{xi_topo}) and~(\ref{g_topo}).

From the condition $M(N_2) \simeq 1$, we can generalize eq.~(\ref{N_2_1}) as
\begin{eqnarray}
N_{2} \simeq \left[N_R(N_2)\right]^3 \times \left\{
           \begin{array}{ll}
              g    &   \quad  ( p < p^*) \\
              G   &   \quad ( p > p^*)
           \end{array}
        \right.   
\label{N_2_p_1}      
\end{eqnarray}
Using the mesh size relation of eq.~(\ref{g_N_R_p}) at $N=N_1$, i.e., $g_{\natural}(N_1) = N_1$ (cf. eq.~(\ref{N_R_N_1_item})), one can explicitly write down the $p$ dependence of $N_1$ as
\begin{eqnarray}
N_{1} \simeq [N_R(N_1)]^2 \times \left\{ 
           \begin{array}{ll}
                g &  \quad ( p < p^*) \\
                G &   \quad ( p > p^*)
           \end{array}
        \right.   
\label{N_1}      
\end{eqnarray}
Then, eq.~(\ref{N_2_p_1}) for $N_2$ can be collectively written as
\begin{eqnarray}
N_{2} \simeq N_1 \times \frac{\left[N_R(N_2) \right]^3}{\left[N_R(N_1) \right]^2} = N_1 {\tilde \phi}_R^{-2} N_R(N_2)
\label{N_2_p_2}
\end{eqnarray}
which is nothing but the last expression in eq.~(\ref{N_2_1}), i.e., again the $p$ dependence is absorbed in $N_1$.

\begin{figure}[h]
\includegraphics[width=0.38\textwidth]{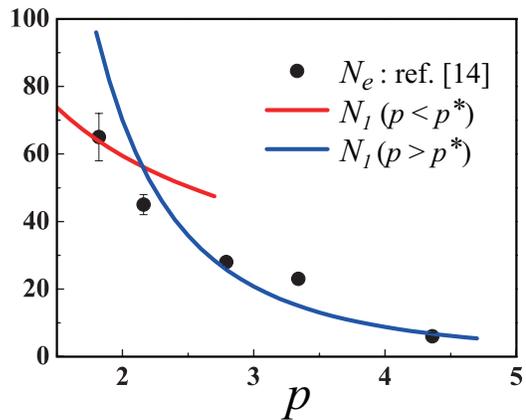}
\caption{(Color online) Dependence of intrinsic topological length scale $N_1$ on the stiffness parameter $p$. Curves are drawn according to eq.~(\ref{N_1}), where numerical coefficients are appropriately adjusted so that the two curves coincide at $p=p^* \simeq 2$. 
Also plotted is the entanglement length $N_e$ in linear chain melts determined by primitive path analysis (points and error bars: from table [1] in ref.~\cite{Everaers_04}).}
\label{Fig6}
\end{figure}
In Fig.~\ref{Fig6}, we plot $N_1$ as a function of $p$ according to eq.~(\ref{N_1}), together with the data of $N_e$ obtained in linear chain melts~\cite{Everaers_04}. It is apparent that these two quantities share the same qualitative trend. Such a coincidence provides a further support for the conjecture $N_1 \simeq N_e$, pointing out an interesting link between the topological constraints in ring polymer systems and the entanglement concept in linear chain systems.

The expression in eq.~(\ref{N_2_p_2}) indicates that the ratio $N_2/N_1$ is a model-independent universal number $\simeq N_R(N_2)^3/N_R(N_1)^2 \sim 50 -100$ independent of $p$. In melts and concentrated solutions, the threshold stiffness $p^*$ is quite small, hence, both onset lengths of the topological effect ($N_1$) and the compact statistics ($N_2$) decrease sharply with the increase in $p$. As a consequence, the region of the crossover regime $N \in (N_1, N_2)$ becomes narrower and the compact statistic regime is reached with shorter rings for a semiflexible ring in agreement with numerical observations~\cite{Halverson_2011,Muller2,Muller3}. Further increase in $p$ may lead to the occurrence of the local orientational ordering~\cite{Muller2}.

\subsection{Comparison with other models}
\label{comparison}
Here we shall look into other models of ring polymer melts in the light of the present theory.

{\it Cates-Deutsch conjecture}:
In their seminal paper, Cates and Deutsch (C-D) argued that unconcatenated rings in the melt may have statistics intermediate between those of compact ($\nu=1/3$) and ideal ($\nu=1/2$) chains~\cite{C-D}. Specifically, they proposed a conjecture on the scaling exponent $\nu=2/5$ by assuming and minimizing the following free energy:
\begin{eqnarray}
\frac{F(R)}{k_BT} \simeq \frac{R^3}{a^3 N} + \frac{Na^2}{R^2}
\label{C-D}
\end{eqnarray}
Their reasoning for the above free energy is as follows.
The first term is the coordination number $N_R$ (with the melt limit $\phi =1$), and they assumed that roughly one degree of freedom is lost for each of the $N_R$ neighbors which the ring is prevented from threading, i.e., the noncatenation constraint.
The second term, on the other hand, represents the entropic penalty of squeezing an ideal chain into a small space of size $R < a N^{1/2}$.
We now argue that (i) the first term (inter-molecular effect) in eq.~(\ref{C-D}) corresponds to the ``dilute'' limit in our theory, and (ii) the second term (intra-molecular effect) is qualitatively incorrect.

(i) First, eq.~(\ref{F_inter}) for $F_{inter}$ can be expanded in a virial series $F_{inter}/k_BT \simeq \phi_R + \phi_R^2/2 + \cdots $. Retaining only a lowest order term (second virial approximation), one finds $F_{inter}/k_BT \simeq R^3 \phi Y/(a^3 N)$, which corresponds to the free energy adopted in the C-D theory aside from a factor $Y$. Indeed, Cates and Deutsch were careful enough to make comment that the estimated free energy $F_{inter}/(k_BT) \sim  N_R$ due to the nonconcatenation constraint may be generally written in the form $ \sim N_R^{\alpha}$ with an unknown exponent $\alpha$.
Our proposal utilizing the van der Waals theory provides a possible (yet nontrivial) way to gather up all the terms, leading to $F_{inter}$ eq.~(\ref{F_inter}). In addition, it turned out that the higher order terms neglected in C-D theory should be important.

(ii) To see the second point, let us remember the discussion in Sec.~\ref{Unknotting_constraint}. Since any ring in melts is squeezed by surrounding rings from all directions, the situation is reminiscent to the chain confined in a closed cavity. As we have already seen, the scaling form in eq.~(\ref{C-D}) is only valid for an ideal chain.
To make the point clear, it is convenient to think about a reference linear chain in the linear chain melt. If, by some mean, we pressurise this reference chain into a small size $R<aN^{1/2}$, the associated free energy is indeed given by the second term in eq.~(\ref{C-D}). However, this does not apply to the ring polymer counterpart because of the topological reason. Indeed, the squeezed ring polymer takes a special entanglement-free conformation called {\it crumpled globule}~\cite{crumpled_globule} (see below), whose statistical properties are rather different from those of a squeezed ideal chain. This is manifested, for example, in the profile of the segment concentration along the cavity axis for the squeezed ring.  While the sinusoidal type function describes that of the ideal linear chain (or ideal phantom ring)~\cite{Grosberg_Khokhlov}, our blob picture in Sec.~\ref{Unknotting_constraint} suggests that the ring polymer develops a bulk plateau profile except for the boundary region of the size $\xi_{\natural}$ (eq.~(\ref{xi_topo})). Such a feature has indeed been observed in numerical simulation~\cite{Flikkema_Brinke}.
Therefore, the main free energy penalty associated with squeezing the ring polymer arises from the unknotting constraint, and this is evaluated as eq.~(\ref{F_intra_0}),~(\ref{F_intra}) or~(\ref{F_intra_2}) as we have argued in Sec.~\ref{Unknotting_constraint}.
In summary, both the inter-molecular and intra-molecular free energies are underestimated in C-D theory, and the exponent $\nu = 2/5$ results from the cancellation of these two errors.

{\it Lattice animal model}:
The idea of reptation in linear chain dynamics was first formulated for a single chain in the fixed obstacles (cross-linked gel)~\cite{deGennes_1971}, and then applied to the rheology of entangle polymer solutions~\cite{Doi_Edwards}. There have been attempts to follow the same procedure in the ring polymer case~\cite{Rubinstein_08,Obukhov,McLeish,Milner_2010}.
A long ring polymer placed in the array of fixed obstacles (without making any concatenation with them) takes a particular conformation known as the lattice animal whose property is well understood~\cite{Zimm_Stockmayer,Lubensky,Daoud_Joanny}.
Indeed, as already stated in Sec.~\ref{conformation}, our results for the compact statistics regime ($N>N_2$) coincides with those obtained for the concentrated solutions of randomly branched polymers, which are continuous analogues of lattice animals.

In melts of ring polymers, there are no fixed obstacles, but each ring feels topological constraints due to the presence of surrounding rings, and one may say that this could be mapped onto the single ring problem in the ``mean-field" array of fixed obstacles with their spacing $\xi_{\natural}$ created by other rings.
There is one important missing ingredient, however, which should be properly taken into account.
Although tempting, the topological effect in the above picture is undertaken only in the inter-molecular interactions. 
In reality, however, the ``mean-field" can self-adjust by weakening the inter-molecular penalty at the expense of the intra-molecular one (see discussion in Sec.~\ref{physical_picture}). The present theory provides a physical picture of how this self-adjustment is achieved, indicating that the topological meshes in individual rings build up along with increasing contact number $N_R$. Hence, the ``mean-field" obstacles should be assumed to originate from both inter- and intra-molecular topological constraints, and the resultant conformation of individual rings is reminiscent to that of crumpled globules~\cite{crumpled_globule} (see below).


{\it Crumpled globule model}:
The concept was originally proposed by Grosberg, et. al. as a long-lived kinetic intermediate on the pathway of the linear chain collapse~\cite{crumpled_globule}, and then, hypothesized as a large scale DNA organization in interphase chromosomes~\cite{crumpled_globule2}. More recently, its relevancy to the conformation of individual rings in their melts has also been discussed~\cite{Vettorel,Halverson_2011}.

To see the point, let us revisit our discussion in Sec.~\ref{Unknotting_constraint}, where we derived the free energy term $F_{intra}$ due to intra-molecular topological constraints in analogy with the confined polymer problem.
When confined in a small closed cavity with the size $D$ smaller than the bulk coil size $R_0 \simeq a N^{\nu_3}$, the polymer takes a compact globule conformation.
For a linear polymer, it can be viewed as a piece of melts, or more precisely semidilute solutions with the mesh size $\xi \simeq a (a/D)^{3\nu_3/(1-3\nu_3)}N^{\nu_3/(1-3\nu_3)}$~\cite{Grosberg_Khokhlov,Sakaue_Raphael}.
It implies that there are many intra-molecular entanglements (self-knotting), and the Flory theorem tells that the conformation of the sub-chain is characterized by the trajectory with the fractal dimension $D_f = 2$ in the scale $\xi < r < D$ because of the screening of excluded-volume interactions.
However, such a self-knotting is forbidden for ring polymers, and upon confinement in a cavity, it follows a self-similar space-filling trajectory with $D_f = 3$, that is the crumpled globule.
Here, all chain parts on scales larger than the topological mesh $\xi_{\natural}$ are mutually segregated in space due to the topological constraint.

Our intra-molecular free energy $F_{intra}$ indeed follows from the crumpled globule concept, which combined with the inter-molecular squeezing effect $F_{inter}$ naturally leads to the conformational properties discussed in Sec.~\ref{conformation}.
But again, there is one important difference compared to the above example of rings under external confinement, where the topological mesh size is assumed to be constant (usually assumed to be $\xi_{\natural} \simeq  \xi (N_e/g)^{1/2}$). The same applies to the problem of the concentration solutions of randomly branched polymers~\cite{Daoud_Joanny} (see Sec.~\ref{conformation}), in which the corresponding scale $\xi_{\natural}$ is set by the degree of the branching $\Lambda$, that is a quenched variable.
In contrast, as explicitly written in eq.~(\ref{R_formula_2}), the topological mesh size $\xi_{\natural}$ in the melts problem is {\it scale dependent}, which should be self-consistently determined by the balance between $F_{inter}$ and $F_{intra}$. This leads to a broad crossover regime with the fictive exponent $\nu_{eff} \simeq 2/5$ during which the topological constraints of the system becomes tighter and tighter, as we have already seen in Sec.~\ref{conformation},~\ref{Large_Scale} and~\ref{physical_picture}.

\section{Summary}
\label{Conclusion}
In this paper, we have sought for a geometrical representation of topological effects in melts and concentrated solutions of ring polymers.
Our central idea lies in the introduction of the topological volume fraction, through which the noncatenation topological constraint is implemented via the effective excluded-volume effect.
From this, we have constructed a mean-field theory, which provides us with a self-consistent topological constraint determined by the balance between inter-molecular noncatenation and intra-molecular unknotting requirements.

From the beginning, we have assumed the presence of an intrinsic topological length scale $N_1$, which signals the onset of the topological effect, thus playing a role analogous to the entanglement length $N_e$ in the rheology of linear chain melts. Once $N_1$ is fixed, our theory identifies two additional length scales $g^* = N_1{\tilde \phi}_R^{-2}$ and $N_2= N_1 {\tilde \phi}_R^{-2} N_R(N_2) = g^* N_R(N_2)$. The latter signals the onset of the compact statistic regime, while the former represents the number of segments in the topological mesh in that regime. 

The geometrical interpretation of $N_1$ is provided via the coordination number $N_R(N)$. We have seen that its values $N_R(N_1)$ at $N=N_1$ and $N_R(N_2)$ at $N=N_2$ are indeed geometrical quantities associated with the random packing properties in three-dimensional space. Therefore, the ratios $g^*/N_1 = {\tilde \phi}_R^{-2} = [N_R(N_2)/N_R(N_1)]^2$ and $N_2/N_1 = N_R(N_2)^3/N_R(N_1)^2$ are universal in the sense that they are insensitive to structural details of polymer models. From such a consideration, these quantities are estimated as $N_R(N_1) \sim 6$ and $N_R(N_2) \sim 15$, which then yield $g^* \sim 6 N_1$ and $N_2 \sim 10^2 N_1$. 

We thus find two well-defined regimes (i) the ideal statistic regime $R = \xi (N/g)^{1/2} = a \phi^{-1/8} N^{1/2}$ for short rings ($N < N_1$) and (ii) the compact statistic regime $R = \xi^* (N/g^*)^{1/3}=a \phi^{-1/8} g^{*1/6} N^{1/3}$ for long rings ($N > N_2$), which are connected through a rather wide crossover regime. By comparing the above two expressions for the ring size in respective regimes, one finds that they match at $N=g^*$. However, these regimes (i) and (ii) are restricted in the range $N<N_1=g^* {\tilde \phi}_R^{2}$ and $N>N_2 = g^*N_R(N_2)$, a straightforward consequence of which is an effective exponent $\nu_{eff}$ (if defined) in this crossover regime should lie in the range $\nu_{eff} \in (1/2, 1/3)$.
The closer analysis shows that the ring size obeys $R = \xi_{\natural} (N/g_{\natural})^{1/3}=a \phi^{-1/8} g_{\natural}^{1/6} N^{1/3}$ in this regime, with the scale-dependent mesh size $\xi_{\natural}(N) =a \phi^{-1/8}[g_{\natural}(N)]^{1/2}$. 
Such a feature should be manifested in the intra-molecular correlation (form factor) of individual rings in melts.
The evolution of $\xi_{\natural}(N) = \xi^* \phi_R^{(eq)}$ with $N$ is controlled by the balance between inter-molecular and intra-molecular topological constraints. Its growth rate is indeed slow, limited by the logarithmic divergence in $F_{inter}(\phi_R)$ with $\phi_R \rightarrow 1$, and well fitted by the effective power law $\xi_{\natural} \sim N^{1/5} \Leftrightarrow g_{\natural} \sim N^{2/5}$, yielding $\nu_{eff} \simeq 1/3 + (2/5)(1/6) = 2/5$. Our theory thus claims that the exponent proposed in C-D conjecture is a fictive one. Nevertheless, the presence of the wide crossover range put forward the practical usage of this fictive exponent, based upon which various effective power laws on statics and dynamics in the crossover should be constructed.

We have discussed the underlying physical picture behind the excluded-volume analogy in Sec.~\ref{physical_picture}. Here it should be emphasized that compared to the usual hard sphere systems, the unique property of ring polymer melts show up owing to their great flexibility and internal degrees of freedom, which is incorporated in the present model through the Y-factor. This softness factor is a consequence of the presence of intrinsic topological length scale $N_1$, allowing partial overlapping (interpenetration) of different rings, and provides a leeway to form the topologically active network of meshes, i.e., $\phi_R(N) = N_R(N) Y < 1$ at $N_1 < N < N_2$ (eq.~(\ref{phi_R})). A fictive exponent $\nu_{eff} \simeq 2/5$, which is slightly larger than that for the compact statistics, results from the progressive interpenetration of surrounding rings. From the observation that such a leeway is gradually lost with the increase in $N_R(N)$ towards $N_R(N_2)$, we find that the presence of this broad crossover regime is inextricably linked to the presence of the intrinsic length scale $N_1$.
Beyond the onset of the compact statistics $N > N_2$, each ring resembles a hard ellipsoid in a statistical sense with the coordination number $N_R(N_2)$ equal to that in hard ellipsoid systems, but there should be, of course, substantial size and shape fluctuations, and the rings can be rather intensively deformed by strong external forcing.

The stiffness effect apparent in the semiflexible rings is shown to be absorbed in $N_1$. Here, it was demonstrated that the stiffness dependence of $N_1$ (eq.~(\ref{N_1})) closely resembles that of $N_e$ in the corresponding linear chain melts (Fig.\ref{Fig6}). This result together with our geometrical interpretation of $N_1$ may provide an intriguing route towards a better understanding of the entanglement concept.

In closing, we would like to emphasize that the melt of ring polymers is not a mere theoretical construction. In addition to past experimental attempts~\cite{Rubinstein_08,Takano,Roovers,Arrighi}, recent progresses in synthesis and characterization methods~\cite{Catalyst,Takano_1,Takano_2} are worth noting, which may open up the quest for unique material properties of topological origin~\cite{McLeish}.
Its close connection to the chromosome territories, i.e., a higher-order spatial organization of chromosomes inside the cell nucleus, provides a further motivation for the investigation~\cite{crumpled_globule2,chromosome_territories}.
There are also other related issues such as topological effects in phase separation~\cite{Ring_linear_compatibility}, confined ring polymer melts~\cite{Yeng-Long} and the property of brush composed of ring polymers~\cite{RingBrush}. 
We hope that the present framework and the emergent geometrical picture will improve our understanding of polymeric systems under topological constraints. The elucidation of the dynamics and rheology~\cite{Rubinstein_08,Halverson_2011,Milner_2010} of ring polymer melts from the present viewpoint might be an interesting challenge.

\acknowledgements
T.S thanks A. Takano for discussion on the experimental situation of ring polymer systems including sample preparation, characterization methods and physical measurements. He is also grateful to J. Juul for his kind help in proofreading.



\begin{thebibliography}{40}
\bibitem{PNAS_2006}D.M. Raymer and D.E. Smith, Proc. Natl. Acad. Sci. USA {\bf 104}, 16432 (2007).
\bibitem{deGennes}P.-G. de Gennes, {\it Scaling Concepts in Polymer Physics} (Cornell University Press, Ithaca, 1979).
\bibitem{Doi_Edwards}M. Doi and S.F. Edwards, {\it The Theory of Polymer Dynamics} (Oxford University Press, Oxford, 1986).
\bibitem{Rubinstein_Colby}M. Rubinstein and R.H. Colby, {\it Polymer Physics} (Oxford University Press, Oxford, 2003).
\bibitem{Edwards_1967}S.F. Edwards, Proc. Phys. Soc. {\bf 92}, 9 (1967).
\bibitem{deGennes_1971}P.-G. de Gennes, J. Chem. Phys. {\bf 55}, 572 (1971).
\bibitem{C-D}M.E. Cates and J.M. Deutsch, J. Physique {\bf 47}, 2121 (1986).
\bibitem{McLeish}T. McLeish, Science {\bf 297}, 2005 (2002).
\bibitem{Grosberg_Khokhlov} A. Grosberg and A. Khokhlov, {\it  Statistical Physics of Macromolecules} (AIP, NY, 1994).
\bibitem{Rubinstein_08}M. Kapnistos et. al. Nature Mater. {\bf 7}, 997 (2008).
\bibitem{Vettorel}T. Vettorel, A.Yu. Grosberg and K. Kremer Phys. Biol. {\bf 6}, 025013 (2009).
\bibitem{Halverson_2011}J.D. Halverson et. al. J. Chem. Phys. {\bf 134}, 204904 (2011); ibid {\bf 134}, 204905 (2011).
\bibitem{Sakaue_2011}T. Sakaue, Phys. Rev. Lett. {\bf 106}, 167802 (2011).
\bibitem{Everaers_04}R. Everaers, et. al., Science {\bf 303}, 823 (2004). 
\bibitem{Muller2}M.M\"uller, J.P. Wittmer and M.E. Cates, Phys. Rev. E {\bf 61}, 4078 (2000).
\bibitem{Muller3}M.M\"uller, J.P. Wittmer and J.-L. Barrat, Europhys. Lett. {\bf 52}, 406 (2000).
\bibitem{Suzuki}J. Suzuki, A. Takano, T. Deguchi and Y. Matsushita, J. Chem. Phys., {\bf 131}, 144902 (2009).
\bibitem{Takano}A. Takano, Polym. Prepr. Jpn. {\bf 56}, 2424 (2007).
\bibitem{Frank-Kamenetskii}M.D. Frank-Kamenetskii, A.V. Lukashin and A. V. Vologodskii, Nature {\bf 258}, 398 (1975).
\bibitem{desCloizeaux}J.des Cloizeaux, J. Phys. Lett. {\bf 42}, L433 (1981).
\bibitem{Sakaue_Raphael}T. Sakaue and E. Rapha\"el, Macromolecules {\bf 39}, 2621 (2006).
\bibitem{Flikkema_Brinke}E. Flikkema and G. ten Brinke, J. Chem. Phys. {\bf 113}, 11393 (2000).
\bibitem{Chaikin_06}P. Chaikin, A. Donev, W. Man, F. Stillinger and S. Torquato, Ind. Eng. Chem. Res. {\bf 45}, 6960 (2006).
\bibitem{crumpled_globule}A.Yu. Grosberg, S.K. Nechaev and E.I. Shakhnovich, J. Phys. France {\bf 49}, 2095 (1988).
\bibitem{Pakula}T. Pakula and S. Geyler, Macromolecules {\bf 21}, 1665 (1988).
\bibitem{Muller1}M.M\"uller, J.P. Wittmer and M.E. Cates, Phys. Rev. E {\bf 53}, 5063 (1996).
\bibitem{Brown}S. Brown, G. Szamel, J. Chem. Phys. {\bf 109}, 6184 (1998).
\bibitem{Daoud_Joanny}M. Daoud and J.F. Joanny, J. Physique {\bf 42}, 1359 (1981).
\bibitem{Khokhlov_Nechaev}A.R. Khokhlov and S.K. Nechaev, Phys. Lett. A {\bf 112}, 156 (1985).
\bibitem{Obukhov}S.P. Obukhov, M. Rubinstein and T. Duke, Phys. Rev. Lett. {\bf 73}, 1263 (1994). 
\bibitem{Milner_2010}S.T. Milner and J.D. Newhall, Phys. Rev. Lett. {\bf 105}, 208302 (2010).
\bibitem{Zimm_Stockmayer}B.H. Zimm and W.H. Stockmayer, J. Chem. Phys. {\bf 17}, 1301 (1949).
\bibitem{Lubensky}T.C. Lubensky and J. Isaacson, Phys. Rev. A {\bf 20}, 2130 (1979).
\bibitem{Schaefer}D.W. Schaefer, J.F. Joanny and P. Pincus, Macromolecules {\bf 13}, 1280 (1980). 
\bibitem{Sakaue_2007}T. Sakaue, Macromolecules {\bf 40}, 5206 (2007).
\bibitem{crumpled_globule2}A.Yu. Grosberg, Y. Rabin, S. Havlin and A. Neer, Europhys. Lett. {\bf 23}, 373 (1993).








\bibitem{Roovers}J. Roovers, Macromolecules {\bf 21}, 1517 (1998).
\bibitem{Arrighi}V. Arrighi, et. al., Macromolecules {\bf 37}, 8057 (2004).

\bibitem{Catalyst}C.W. Bielawski, D. Benitez, R.H. Grubbs, Seience {\bf 297}, 2041 (2002).
\bibitem{Takano_1}A. Takano et. al., Macromolecules {\bf 40}, 679 (2007).
\bibitem{Takano_2}Y. Ohta, Y. Kushida, Y. Matsushita and A. Takano, Polymer {\bf 50}, 1297 (2009).
\bibitem{chromosome_territories}E. Lieberman-Aiden, et. al., Science {\bf 326}, 289 (2009).
\bibitem{Ring_linear_compatibility}A.R. Khokhlov and S.K. Nechaev, J. Phys. II France {\bf 6}, 1547 (1996).
\bibitem{Yeng-Long} Private communication with Y.-L. Chen.
\bibitem{RingBrush}D. Reith, A. Milchev, P. Virnau and K. Binder, Europhys. Lett. {\bf 95}, 28003 (2011).
























\end{thebibliography}
\end{document}